\documentclass[twocolumn,showpacs,preprintnumbers,amsmath,amssymb,superscriptaddress, natbib]{revtex4-1}
\usepackage{hyperref}

\usepackage{graphicx}
\usepackage{epstopdf}
\DeclareGraphicsExtensions{.pdf,.jpg,.png,.eps}
\usepackage{color}
\usepackage{color}
\usepackage{epstopdf}
\usepackage{times}
\usepackage[symbol]{footmisc}

\newcommand{\tfco}{TbFe$_{0.5}$Cr$_{0.5}$O$_3$}
\newcommand{\tn}{$T_N$}
\newcommand{\tsr}{$T_\mathrm{SR}$}

\begin{document}

\title{ Reentrant spin reorientation transition and Griffiths$\textendash$like phase in antiferromagnetic \tfco}

\author{Bhawana Mali}
\thanks{Corresponding author email id: bhawana@iisc.ac.in}
\affiliation{Department of Physics, Indian Institute of Science, Bangalore 560012, India}
\author{Harikrishnan S. Nair}
\affiliation{Department of Physics, 500 W. University Ave, The University of Texas at El Paso, TX 79968, USA}
\author{ T. W. Heitmann}
\affiliation{University of Missouri Research Reactor, University of Missouri, Columbia, MO 65211, USA}
\author{ Hariharan Nhalil}
\thanks{Present address: Department of Physics, Bar-Ilan University Ramat-Gan, Israel}

\affiliation{Department of Physics, Indian Institute of Science, Bangalore 560012, India}
\author{Daniel Antonio}
\affiliation{Idaho National Laboratory, Idaho Falls, ID 83415, USA}
\author{Krzysztof Gofryk}
\affiliation{Idaho National Laboratory, Idaho Falls, ID 83415, USA}
\author{Shalika Ram Bhandari}
\affiliation{Central Department of Physics, Tribhuvan University, Kirtipur, 44613, Kathmandu, Nepal}
\affiliation{IFW Dresden, Helmholtzstr. 20, D-01069, Dresden, Germany}
\author{Madhav Prasad Ghimire}
\affiliation{Central Department of Physics, Tribhuvan University, Kirtipur, 44613, Kathmandu, Nepal}
\affiliation{IFW Dresden, Helmholtzstr. 20, D-01069, Dresden, Germany}
\author{Suja Elizabeth}
\affiliation{Department of Physics, Indian Institute of Science, Bangalore 560012, India}

\date{\today}
\begin{abstract}
The perovskite \tfco\ shows two anomalies in the magnetic
susceptibility at \tn\ = 257~K and 
\tsr\ = 190~K which are respectively, the antiferromagnetic 
and spin reorientation transition that occur in the Fe/Cr sublattice. 
Analysis of the magnetic susceptibility reveals signatures of
Griffiths$\textendash$like phase in this compound: 
the negative deviation from ideal Curie$\textendash$Weiss law and in 
less$\textendash$than$\textendash$unity power$\textendash$law 
susceptibility exponents.
Neutron diffraction analysis confirms that, as the temperature is 
reduced from 350~K, a spin reorientation transition 
from $\Gamma_2$ (F$_x$, C$_y$, G$_z$) to $\Gamma_4$ (G$_x$, A$_y$, F$_z$) 
occurs at \tn\ = 257~K and subsequently, a second spin reorientation takes place 
from $\Gamma_4$ (G$_x$, A$_y$, F$_z$) to $\Gamma_2$ (F$_x$, C$_y$, G$_z$) 
at \tsr\ = 190~K. 
The $\Gamma_2$ (F$_x$, C$_y$, G$_z$) structure is 
stable until 7.7~K where an ordered moment of 
7.74(1)$\mu_\mathrm B$/Fe$^{3+}$(Cr$^{3+}$) 
is obtained from neutron data refinement. 
In addition to the long$\textendash$range order of the magnetic structure, 
indication of diffuse magnetic scattering at 7.7~K is 
evident, thereby lending support to the 
Griffiths$\textendash$like phase observed in susceptibility. 
At 7.7~K, Tb develops a ferromagnetic component along the crystallographic $a$ axis.
Thermal conductivity, and spin$\textendash$phonon coupling of \tfco\ through 
Raman spectroscopy are studied in the present work.
The magnetic anomalies at \tn\ and \tsr\ do not reflect in the thermal conductivity data of \tfco;
however, it is noticeable that the application of 9~T magnetic field has no effect on 
the thermal conductivity. 
The  \tn\ and \tsr\ are revealed in the temperature$\textendash$dependence of 
full$\textendash$width$\textendash$at$\textendash$half$\textendash$maximum 
curves obtained from Raman intensities.
An antiferromagnetic structure with ($\uparrow \downarrow \uparrow \downarrow$)
arrangement of Fe/Cr spins is found in the ground state through first$\textendash$principles 
energy calculations which supports the experimental magnetic structure at 7.7~K.
The spin$\textendash$resolved total and partial density of states are determined showing that
\tfco\ is insulating with a band gap of $\sim 0.12$ (2.4)~eV within GGA (GGA+$U$) functionals
\end{abstract}
\maketitle

\section{Introduction}
\indent
Rare earth orthoferrites and orthochromites with the general formula $RM$O$_3$, 
where $R$ = rare earth or yttrium and $M$ = Fe or Cr, crystallize in the 
perovskite structure (usually $Pbnm$ space group) 
with orthorhombic distortion and an antiferromagnetic ground state~\cite{doi:10.1063/1.1657530}. 
Rare earth orthoferrites possess  a complex spin structure and have drawn considerable attention 
due to their unique physical properties~\cite{doi:10.1063/1.1657530} and potential 
applications such as ultrafast magneto$\textendash$optical recording~\cite{PhysRevB.74.060403}, 
laser$\textendash$induced thermal spin reorientation~\cite{Kimel2004}, precision excitation 
induced by terahertz pulses~\cite{doi:10.1063/1.4818135}, inertia$\textendash$driven 
spin switching~\cite{Kimel2009}, and magnetism$\textendash$induced multiferroicity~\cite{Tokunaga2009}. 
Most orthoferrites are G$\textendash$type canted antiferromagnets with a weak ferromagnetic 
component due to Dzyaloshinskii$\textendash$Moriya (DM) interaction 
and show temperature$\textendash$induced spin reorientation (SR) from one magnetic 
symmetry to another. 
In $R$FeO$_3$, exchange interactions between Fe$^{3+}$$\textendash$ Fe$^{3+}$, 
$R^{3+}$$\textendash$Fe$^{3+}$ and $R^{3+}$$\textendash$$R^{3+}$ play an 
important role in determining complex magnetic structures. 
Isotropic Fe$^{3+}$$\textendash$Fe$^{3+}$ exchange interaction determines the 
magnetic structure of Fe$^{3+}$ spins below the antiferromagnetic ordering temperature. 
Exchange field due to Fe$^{3+}$ moment polarizes the $R^{3+}$ spins of the $R$ sublattice
and the Fe$^{3+}$$\textendash$$R^{3+}$ interaction, in turn,  generates 
effective fields on Fe$^{3+}$ spins which undergo spin reorientation transition and align 
perpendicular to the $R^{3+}$ spins. 
The spin reorientation transition might be continuous or abrupt depending on the $R$ element~\cite{YAMAGUCHI1974479}. \\
\indent
 In TbFeO$_3$, an unusual incommensurate magnetic phase was 
discovered~\cite{Artyukhin2012} and it was shown that the exchange of spin waves between 
extended topological defects could result in novel magnetic phases which draws parallels with the 
Yukawa forces that mediate between protons and neutrons in a nucleus. 
The Fe$^{3+}$ moments in TbFeO$_3$ exhibit G$_x$A$_y$F$_z$ ($Pb'n'm$) 
spin configuration at room temperature~\cite{refId0,BERTAUT1967293, 0295-5075-30-4-007} 
which is accompanied by a spin reorientation to 
F$_x$C$_y$G$_z$ ($Pbn'm'$). At 3~K, another spin reorientation occurs to revert to the 
G$_x$A$_y$F$_z$ ($Pb'n'm$) structure. In recent years, a variety of interesting 
properties were achieved by substituting Fe ion by different transition metal 
ions~\cite{TAGUCHI1997108,DAHMANI2003912}. 
According to Goodenough$\textendash$Kanamori rules~\cite{PhysRev.100.564}, Cr$^{3+}$ is a good 
choice to pair with Fe$^{3+}$ to tune superior magnetic properties due to superexchange 
interaction between empty e$_g$ orbital of Cr$^{3+}$ and half filled e$_{g}$ orbital of Fe$^{3+}$ ions. \\
\indent
In TbCrO$_3$, the exchange coupling between the nearest neighbour Cr$^{3+}$ 
is predominantly antiferromagnetic and the Cr$^{3+}$ spins order spontaneously 
at \tn\ = 167~K~\cite{PhysRevB.13.3012}. 
Below this temperature, it exhibits weak ferromagnetism resulting from the 
canting of Cr$^{3+}$ magnetic moments. 
In TbCrO$_3$ the Cr$^{3+}$ spin structure is G$_z$F$_x$ below \tn\ 
and belongs to $\Gamma_2$ configuration which implies that the 
weak ferromagnet component of the Cr$^{3+}$ moments orient along the $a$ axis~\cite{BERTAUT19672143,1065951}. 

Tb$^{3+}$ spins order antiferromagnetically at 3.05~K~\cite{BERTAUT19672143}
below which temperature the Tb$^{3+}$ spin system exhibits an A$_x$G$_y$ structure. 
In the temperature range, 3.05~K $<$ T $<$ \tn\, the Tb$^{3+}$ spin system has F$_x$C$_y$ 
structure which belongs to $\Gamma_4$ representation and is coupled to the ordered 
Cr$^{3+}$ spin system~\cite{BERTAUT19672143}. Spin reorientation, magnetization reversal and weak ferromagnetism is seen in compounds like TbFe$_{0.5}$Mn$_{0.5}$O$_3$~\cite{doi:10.1063/1.4919660}. And 
reentrant spin reorientation transition has been seen in compounds like 
TbFe$_{0.75}$Mn$_{0.25}$O$_3$~\cite{Fang2016} which undergo $
\Gamma_4$ to $\Gamma_1$ transition and then, $\Gamma_1$ to $\Gamma_4$ transition.
Recently, the magnetic structures and spin reorientation transitions of the 
mixed orthochromite$\textendash$orthoferrite perovskites RFe$_{0.5}$Cr$_{0.5}$O$_3$, 
where R = Tb, Dy, Ho, Er have been reported~\cite{PhysRevB.98.134417} using neutron diffraction.\\
\indent
In the present paper, we report a detailed study of magnetic phase transitions and 
magnetic structure of \tfco\ through magnetization, neutron powder diffraction, 
Raman scattering and thermal conductivity studies in conjunction with 
density functional theory calculations. Our results support reentrant spin reorientation transition and a Griffiths phase$\textendash$like features in the title compound. Our neutron scattering study also gives indication of diffuse magnetic component present below the T$_N$.

\section{Experimental methods}
\indent Polycrystalline \tfco\ was prepared by standard solid state reaction using high purity ($\ge$3N) Tb$_4$O$_7$, Fe$_2$O$_3$ and Cr$_2$O$_3$ in 
stoichiometric amounts. 
The starting materials were thoroughly mixed and sintered at 
$1200^{\circ}$C for 48~h with two times intermediate grinding. 
The phase purity of sintered sample was verified by taking 
powder X$\textendash$ray diffractograms (PXRD) using
Rigaku Smartlab X$\textendash$ray diffractometer with Cu K$_\alpha$ radiation ($\lambda$ = 1.548~{\AA}). Oxidation states of Fe and Cr ions were 
determined using X$\textendash$ray photoelectron 
spectroscopy (XPS) in a AXIS Ultra spectrometer and the data was 
analyzed using the CASA XPS spectroscopy software~\cite{casaxps}. 
The chemical composition analysis of powder samples were performed 
using JEOL$\textendash$JXA$\textendash$8530F electron probe micro analyzer (EPMA)  which 
yielded the Fe:Cr atomic ratio as 0.48:0.50 (Fe/Cr = 0.96).   
Temperature dependent DC magnetization measurements were performed on 
sintered pellets using a commercial magnetic property measurement system 
(MPMS, Quantum Design) in the temperature range of 5~K $\leq T \leq$ 400~K at 
100~Oe and 500~Oe in both zero$\textendash$field cooled (ZFC) and 
field cooled (FC) protocols. 
Additionally, high temperature magnetic susceptibility 
was recorded up to 800~K in the high temperature
VSM oven option provided with the physical property measurement system (PPMS). 
The thermal conductivity of a parallelopiped sample 
of \tfco\ was measured in the temperature 
range 2~K$\textendash$300~K in 0~T and 9~T magnetic field 
using a commercial physical property measurement system 
(PPMS).\\
\indent 
To investigate nuclear and magnetic structure of \tfco, 
neutron powder diffraction experiments were 
performed at University of Missouri Research Reactor (MURR) 
using the neutron powder diffractometer, PSD. 
Neutron powder diffraction patterns of 2~g powder sample were 
collected at 350~K, 300~K, 215~K, 100~K and 7.7~K using
neutrons of wavelength 1.485~{\AA}. 
The neutron diffraction data were analyzed using Fullprof suite 
of programs~\cite{FullProf_carvajal} employing the 
Rietveld method~\cite{rietveld1969profile}. 
Magnetic representations belonging to the $Pbnm$ symmetry were 
determined using the software SARA$h$~\cite{wills_sarah} 
and the corresponding magnetic structure was refined using Fullprof.  
Raman spectra was recorded from 110~K to 300~K 
temperature range in the backscattering geometry by 
using a HORIBA JOBIN$\textendash$YVON spectrometer with 633~nm laser as an excitation source. 
Low temperature was maintained by closed cycle 
He$\textendash$cryostat attached to spectrometer.

\section{Computational Details}
The electronic and magnetic structure calculations were performed by means of density$\textendash$functional theory
(DFT) approach using the full$\textendash$potential linearized augmented plane wave plus local orbital method as
implemented in the WIEN2k code~\cite{blaha2001wien2k}. 
The non$\textendash$overlapping muffin$\textendash$tin sphere radii (R$_{MT}$) of 2.35, 2.0, 1.96, and 1.72 
Bohr were used for Tb, Fe, Cr, and O respectively. 
The linear tetrahedron method with 500 $k$ points was employed for
the reciprocal$\textendash$space integrations in the whole Brillouin zone (BZ) that 
corresponds to 216 $k$$\textendash$points within the irreducible BZ.
For the calculations, the standard generalized$\textendash$gradient approximation (GGA)
in the parameterization of Perdew, Burke, and Ernzerhof
(PBE$\textendash$96)  was used~\cite{perdew1996generalized}.
In order to consider the strong correlation effects, GGA+$U$ functional
with double$\textendash$counting corrections according to Anisimov {\em et al.}~\cite{anisimov1997first}
was used.
The chosen values of $U$ were 6 eV for Tb$\textendash$4$f$, 5 eV for Fe$\textendash$3$d$, and 3 eV
for Cr$\textendash$3$d$ states, which are comparable to the values found in literature~\cite{ghimire2016possible,ghimire2016compensated,ghimire2010study,yuan2015high,feng2016b,Feng_19}.
Calculations were performed using the lattice parameters obtained from neutron 
diffraction data at 7.7~K (see Table~\ref{tab:str}).
The energy and charge convergence was set to 10$^{-6}$ Ry and 10$^{-4}$ of an 
electron, respectively, for self$\textendash$consistent calculations.
To obtain the magnetic ground states, we have considered five magnetic 
configurations by computing their total energies. 
They are ferromagnetic (FM$\textendash$$\uparrow\uparrow\uparrow\uparrow$), 
two antiferromagnetic (AFM1$\textendash$$\uparrow\downarrow\uparrow \downarrow$ and 
AFM2$\textendash$$\uparrow\downarrow\downarrow\uparrow$) and 
two ferrimagnetic (FIM1$\textendash$$\uparrow\uparrow\downarrow\downarrow$ and 
FIM2$\textendash$$\uparrow\uparrow\uparrow\downarrow$). 
Here, the spin arrangements for two inequivalent atoms each of Fe and Cr atoms are 
arranged as Fe1, Fe2, Cr1 and Cr2, respectively. 

\section{Results and discussion}
\begin{figure}[!t]
\centering\includegraphics[trim={0, 0, 0, 0}, scale=0.32]{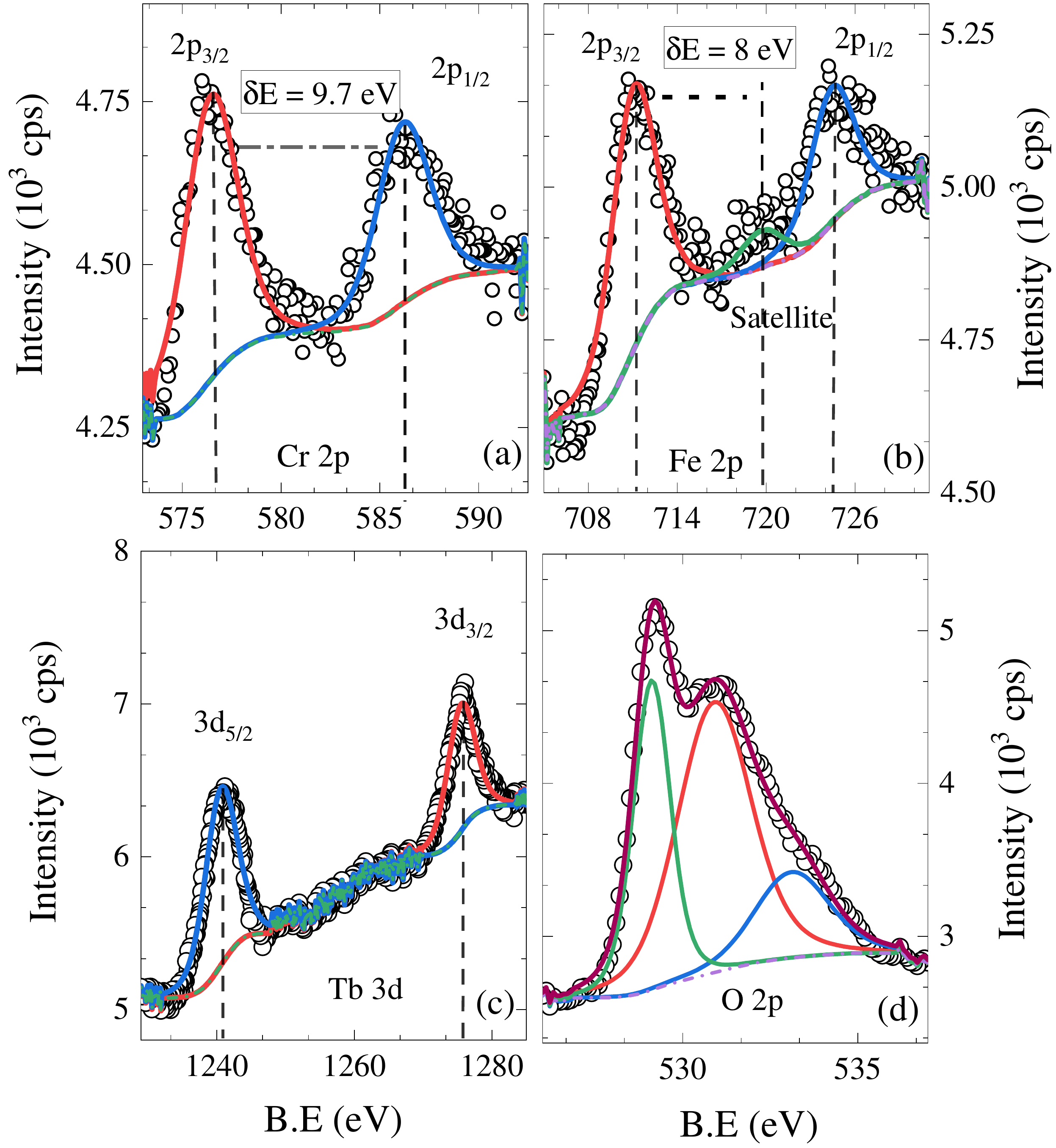}
\caption{\label{fig:xps-fitting} X$\textendash$ray photoelectron spectra of 
	(a) Cr 2$p$, (b) Fe 2$p$, (c) Tb 3$d$ and (d) O 2$p$ are shown 
	in open circles. Solid lines are fitted peaks, deconvoluted 
	components and background, respectively, in each graph. 
	Oxidation state of 3+ is inferred for Fe, Cr and Tb from this data.}
\end{figure}
\subsection{X$\textendash$ray photoelectron spectroscopy}
\indent Core level X$\textendash$ray photoelectron spectroscopy measurements at room 
temperature using Al K$_\alpha$ X$\textendash$ray source was performed to determine the 
valence states of cations in \tfco. 
Figure~\ref{fig:xps-fitting} shows the experimental intensities along with the peak fit obtained 
using the CASA XPS software. 
The core level binding energy was calibrated with carbon (B.E. = 284.8~eV). 
The Cr $2p_{3/2}$ peak at 576.5~eV is close to the binding energy 
of Cr$_2$O$_3$ (576~eV)~\cite{DT9730001675}. 
However, in oxides, $2p_{3/2}$ peak of Fe$^{2+}$ and Fe$^{3+}$ appear 
around the binding energy values of 710.3~eV 
and 711.4~eV, respectively~\cite{DT9740001525}. 
In \tfco, the peak at 711~eV is close to the binding energy value of Fe$^{3+}$. 
Additionally, a satellite peak at 8~eV above the Fe 2$p_{3/2}$ confirms Fe$^{3+}$ state 
(Fe$^{2+}$ gives a satellite peak at 6~eV above the main Fe 2$p_{3/2}$ peak). 
The XPS spectrum of Tb 3$d_{5/2}$ yields a peak at 1240.8~eV which is very close to that 
of Tb$_2$O$_3$ peak (1241.2~eV)~\cite{10.2307/79230}. 
Our XPS results thus indicate 3+ oxidation state in Tb, Fe and Cr.
\begin{figure*}
\centering \includegraphics[trim={0, 0, 0, 0}, scale=0.45]{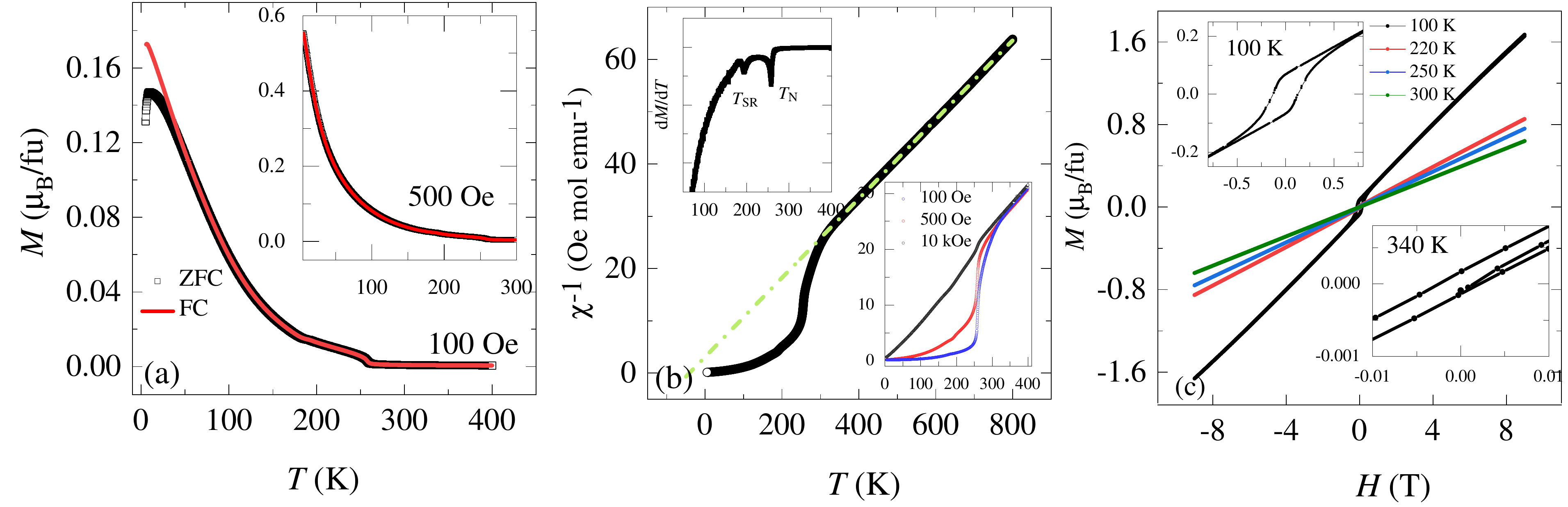}
\caption{\label{fig:mt} (a) Magnetization as a function of temperature showing a bifurcation for ZFC and 
	FC curves and the transitions at \tn$\approx$ 257~K and \tsr$\approx$ 190~K. 
	The inset shows the magnetization at 500~Oe. 
	(b) $\chi^{-1} (T)$ at 500~Oe along with Curie$\textendash$Weiss fit (dashed line). 
	The upper inset shows the derivative $dM$/$dT$ where anomalies at \tn\ and \tsr\ are clear.  
	The lower inset shows the inverse magnetic susceptibility at 100~Oe, 500~Oe 
	and 10~kOe which shows that the negative curvature vanishes at higher fields. 
	(c) The magnetization isotherms, $M (H)$, at 100~K, 220~K, 250~K and 300~K. 
	A weak hysteresis that develops below the \tsr\ is shown in the upper inset ($T$ = 100~K) and the 
	lower inset shows a magnified view of a magnetization isotherm  above the \tn\ ($T$ = 340~K).}
\end{figure*}

\subsection{Magnetic properties: Spin reorientation and Griffiths$\textendash$like phase}
\label{mag}
\begin{figure}[!b]
\centering\includegraphics[trim={1cm, 0, 0, 0}, scale=0.27]{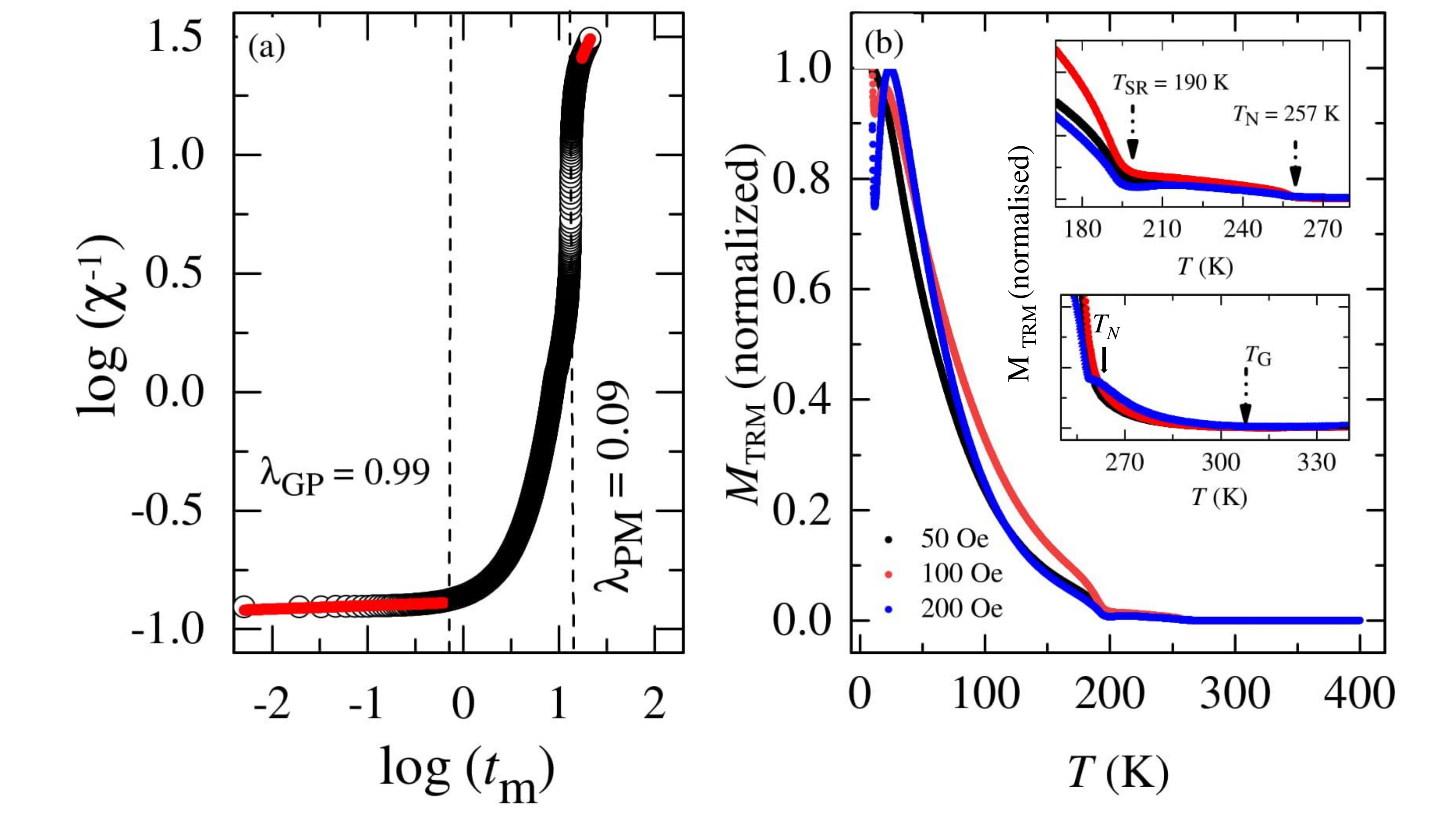}
\caption{\label{fig:TRM} (a) Shows the power law fits (red lines) to 
$\chi^{-1}(T)$ at 100~Oe; plotted in a log$\textendash$log scale.
The reduced$\textendash$temperature is $t_m$ = ($T$/$T^R_c$$\textendash$ 1).
(b) Thermoremanant magnetization, $M_{TRM}$, measured at 50~Oe, 100~Oe 
and 200~Oe cooling fields, showing the onset of spin reorientation transition 
$T_{SR}$ and the antiferromagnetic $T_N$. 
The lower and upper insets show magnified regions neat the \tn\ and \tsr,\ respectively.}
\end{figure}

\indent 
Figure~\ref{fig:mt} (a) shows the temperature dependent magnetization, $M (T)$, of 
\tfco\ under ZFC and FC protocol at 100~Oe and 500~Oe external magnetic field.
Two anomalies are seen in the $M (T)$ curve at $\approx$ 257~K and at 190~K. 
A bifurcation of the ZFC and FC curves is seen below $\approx$15~K. 
With the application of of 500~Oe, the bifurcation vanishes (see the inset of Figure~\ref{fig:mt}(a)).
The magnetic phase transition temperatures of \tfco\ are determined by plotting 
$dM/dT$ vs $T$ as shown in top inset of Figure~\ref{fig:mt} (b). 
\tsr = 190~K and \tn = 257~K are identified in this manner. 
The temperature dependent inverse magnetic susceptibility, 
$\chi^{-1}(T)$,  of \tfco\ at 500~Oe is plotted in the main panel of Figure~\ref{fig:mt} (b) 
up to 800~K along with a curve fit using the Curie$\textendash$Weiss (CW) law (dashed line), 
$\chi^{-1}$ = $(T \textendash \theta)/C$,
where, $C$ = $N_{A} \mu_\mathrm{eff}^2/3k_\mathrm B$ is the Curie 
constant, $N_A$ is the Avogadro's number, $\mu_\mathrm{eff}$ is the 
effective magnetic moment, $k_\mathrm B$ is the Boltzmann constant 
and $\theta$ is the Curie$\textendash$Weiss temperature~\cite{kittel_book}. 
The Curie Weiss analysis  yields an effective magnetic moment of 
$\mu_\mathrm{eff}$ = 10.3(2)~$\mu_\mathrm B$. 
Assuming 3+ oxidation state for Tb, Fe and Cr, as determined from the XPS 
analysis, the theoritically calculated magnetic moment, $\mu_\mathrm{theory}$ in the 
paramagnetic region using the relation, $\mu_\mathrm{theory} = \sqrt{\mu_\mathrm{Tb}^2(\mathrm{Tb}^{3+})+0.5\times\mu_\mathrm{Cr}^2(\mathrm{Cr}^{3+})+0.5\times\mu_\mathrm{Fe}^2(\mathrm{Fe}^{3+})}$ and considering high spin state of Tb$^{3+}$ ($\mu_\mathrm{Tb}$ = 9.7~$\mu_\mathrm B$), Fe$^{3+}$ ($\mu_\mathrm{Fe}$ = 5.9~$\mu_\mathrm B$) and Cr$^{3+}$ ($\mu_\mathrm{Cr}$ = 3.9~$\mu_\mathrm B$) is 10.9~$\mu_\mathrm B$. 
\\
\indent
 A downward deviation of inverse magnetic susceptibility values from the ideal Curie$\textendash$Weiss law 
is a signature of Griffith's phase (GP)~\cite{griffiths1969nonanalytic,nair2011griffiths, chakraborty2016disordered}.  The 
characteristic temperature at which the inverse susceptibility deviates from 
the CW behavior is known as the Griffiths temperature, $T_G$. 
It is clear from the main panel of Figure~\ref{fig:mt} (b) that the inverse susceptibility deviates from CW 
law above \tn\ at $T_G \approx$ 320~K. The downturn softens with increase 
in applied magnetic field as shown in the lower inset of Figure~\ref{fig:mt} (b), this 
is a signature of the presence of Griffiths$\textendash$like phase in the compound. \\
\indent
Figure~\ref{fig:mt} (c) shows the magnetization isotherms of \tfco\ 
at 100~K, 220~K, 250~K, 300~K and 340~K measured upto $\pm$9~T
which do not reveal strong ferromagnetic features. 
However, at 100~K, an opening of the magnetic hysteresis loop is 
observed at low applied field values (upper left inset of Figure~\ref{fig:mt} (c)). 
A magnified view of the isotherm at 340~K ($>$ \tn) shown in lower inset of Figure~\ref{fig:mt} (c) reveals
weak hysteresis indicating the presence of short range magnetism above \tn. 
In section~\ref{npd} we present experimental evidences of spin fluctuation above \tn\ in \tfco.
\\
\indent 
We noted earlier that the downturn softening of the 
downturn in $\chi^{-1} (T)$ with an increase in applied field, supports GP scenario~\cite{PhysRevB.81.024431,0953-8984-28-35-35LT02,PhysRevLett.96.167201}. 
The suppression of downturn in magnetic susceptibility 
at high magnetic fields is due to the rising paramagnetic background which 
masks the ferromagnetic signal.
Griffiths phase consists of finite size ferromagnetic (FM) clusters in a
paramagnetic matrix well above the transition temperature in which 
the spins are ferromagnetically correlated within those clusters. 
However, the magnetic system as a whole does not have long$\textendash$range ordering in GP and 
thus no spontaneous magnetization will appear. 
In GP, the FM clusters will appear with variable sizes having
local ferromagnetic ordering due to which magnetization becomes 
non$\textendash$analytic; in the low$\textendash$field region, the magnetic susceptibility will 
follow a power law behavior~\cite{PhysRevLett.23.17,0953-8984-28-35-35LT02,PhysRevB.81.024431} 
given by, 
$\chi^{-1} \propto$ $(T \textendash T^R_c)^{1-\lambda}$,
where $T^R_c$ is the critical temperature of the FM clusters. Here 
susceptibility tend to deviate from CW law and $\lambda$ ( $0 \le \lambda \le 1$) 
is the exponent which signifies the deviation from 
CW behavior due to formation of magnetic clusters in the PM state 
above the transition temperature. 
A power law fit using the above$\textendash$mentioned equation was
administered on the magnetic susceptibility of \tfco\ as log($\chi$$^{-1}$) versus 
log($T$/$T^R_c \textendash$ 1) in both PM and GP regions as shown 
in Figure~\ref{fig:TRM} (a). Since the value of $\lambda$ is highly 
sensitive to $T^R_c$, we have proceeded to estimate the value of
$T^R_c$ accurately~\cite{0953-8984-28-35-35LT02,doi:10.1063/1.3335895}. 
The critical temperature of ferromagnetic clusters, $T^R_c$, is always greater than 
the transition temperature, so we first estimated the value of 
$T^R_c$ in the purely paramagnetic region. This yields a value of 18~K 
which was later used in the curve$\textendash$fitting for the Griffiths phase 
regime to obtain $\lambda$ = 0.99. 
In the high temperature region, we obtained $\lambda$ as 0.09 which signifies that the system 
is in paramagnetic phase, following the CW behavior. 
These values for $\lambda$ are consistent with the GP model signifying
a Griffiths singularity in \tfco. \\
Griffiths phase features are experimentally observed in
strongly correlated systems like the layered iron pnictides~\cite{inosov2013possible}
and in geometrically frustrated antiferromagnets~\cite{PhysRevB.95.054401}.
In the case of the former, it is formed by randomly introduced
localized magnetic impurities which form above the quantum critical point
associated with the suppression of the stripe$\textendash$antiferromagnetic spin density
wave order. 
In the latter, the Griffiths phase was observed to be robust against the oxygen 
nonstoichiometry which influenced the antiferromagnetic ordering.
Since the total magnetic susceptibility in the Griffiths phase region
contains contributions from both paramagnetic as well 
as short$\textendash$range correlated regions, the downturn observed in the
inverse magnetic susceptibility from ideal CW law is not expected to be
sharp in the case of antiferromagnetically correlated regions.
In $R$FeO$_3$, five outer shell electrons of the Fe$^{3+}$ ion are in 
half$\textendash$filled $e_g$ ($\sigma$$\textendash$bond component) and $t_{2g}$ ($\pi$$\textendash$bond component) 
orbitals resulting in superexchange interactions that are antiferromagnetic. 
In the case of Cr$^{3+}$ ions, superexchange interactions in the 
half$\textendash$filled $t^3 \textendash$O$\textendash t^3$ induce antiferromagnetism. 
Since Fe$^{3+}$ and Cr$^{3+}$ are randomly distributed in the lattice of \tfco, 
it results in stabilization of both ferromagnetic and antiferromagnetic couplings. 
\\
\indent
To confirm the GP$\textendash$like scenario in antiferromagnetic \tfco, 
we employed thermoremanant magnetization protocol to
measure magnetization ($M_{TRM}$), which has been used earlier to study
spin glasses~\cite{PhysRevB.63.092401}. 
This protocol involves cooling the sample from well above the 
magnetic transition temperature in the presence 
of a magnetic field. The field is then switched off below $T_C$, and the 
magnetization measured upon warming in zero field condition. 
The thermoremanent magnetization $M_{TRM}$ will exhibit a sharp upturn 
at the transition temperature. 
In the present case of \tfco, this protocol was repeated for three different 
cooling fields, 50~Oe, 100~Oe and 200~Oe. 
The zero$\textendash$field measurements performed here have the advantage that
the contributions from the paramagnetic susceptibility are suppressed
compared to an in$\textendash$field measurement.
Figure~\ref{fig:TRM} (b) shows $M_{TRM}$(T) measured at the 
different cooling fields. 
A clear signature of GP$\textendash$like phase is seen in the form of an upturn in 
magnetization at temperature well 
above $T_N$. 
The $T_G$ obtained from thermoremanent measurement is 315~K which is close to the value of 320~K estimated from magnetic susceptibility earlier .

\subsection{Neutron diffraction: Reentrant spin reorientation and short$\textendash$range spin correlations}
\label{npd}
The macroscopic magnetic measurements explicitly suggest the antiferromagnetic
ordering at \tn, the possibility of a spin reorientation transition at \tsr\ and the presence of
Griffiths$\textendash$like phase in \tfco. We now proceed to investigate the magnetic structure of \tfco\ in detail
so as to understand the spin reorientation process and to ascertain the magnetic structures
above and below the \tsr. 
For this purpose, neutron diffraction experiments were carried out 
on powder samples of \tfco\ at various temperatures in the 
range of 7.7~K to 350~K. 
\\
\begin{figure}[!b]
\centering
\includegraphics[scale=0.25]{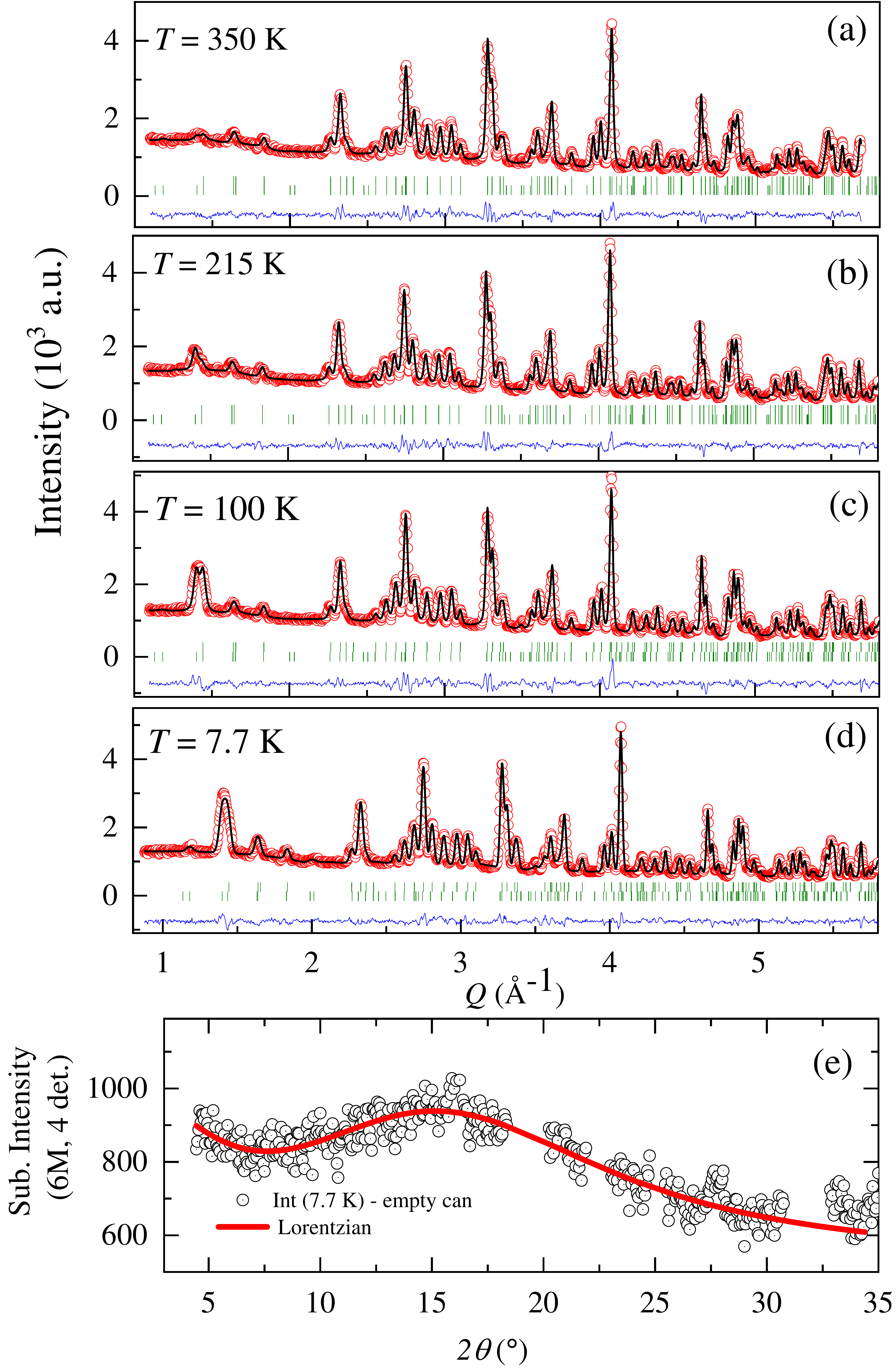}\\
\caption{\label{fig:neutron} (a-d) Rietveld refinement of the neutron powder diffraction 
patterns of \tfco\ at 350~K, 215~K, 100~K, and 7.7~K using $Pbnm$ space group model. There is a weak magnetic
contribution even at 350~K which is above the \tn, observed in magnetometry. (e) Diffuse scattering intensity at 7.7~K 
along with the curve fit (solid line) using a Lorentzian function.}
\end{figure}
\indent
The experimental neutron diffraction patterns at 350~K, 215~K, 100~K and 
7.7~K are shown in Figures~\ref{fig:neutron} (a-d) (red circles). Orthoferrites adopt 
orthorhombic structure as observed in a variety of 
$R$Fe$_{0.5}$Cr$_{0.5}$O$_3$~\cite{bolletta2018spin}.
For $R$ = Tb, Dy, Ho and Er, the crystal structure belongs to a 
distorted perovskite type in the space group $Pbnm$ and
ordered antiferromagnetically below 270~K in $G_x$ configuration 
compatible with the $\Gamma_4$ representation.
The above mentioned compounds exhibited a spin reorientation transition from $G_x$ ($\Gamma_4$) to
$G_z$ ($\Gamma_2$).
In present case of \tfco, refinement of the neutron diffraction data at 350~K
was first performed using purely nuclear space groups, $P2_1/n$ and $Pbnm$.  
The perovskite structure in which cations order crystallographically 
may adopt a doubled unit cell with monoclinic $P2_1/n$ space group~\cite{anderson1993b}.
In \tfco, Rietveld analysis of the diffraction data at 350~K resulted in a reasonably
good fit assuming $Pbnm$ space group, however, the intensity of the nuclear Bragg peak position
(101) was not fully accounted for. 
Even at  350~K, appreciable contribution from magnetic 
scattering towards the total scattered intensity was observed. \\
\indent
In order to determine the magnetic structure, we scrutinized the symmetry$\textendash$allowed magnetic 
structures for this class of compounds. 
It can be seen that for $R$FeO$_3$ compounds in $Pbnm$ space group, 
there exists eight irreducible representations
$\Gamma_1$ through $\Gamma_8$~\cite{doi:10.1063/1.1657530}. 
For the $4b$ Wyckoff position, the configurations  $\Gamma_5$ to $\Gamma_8$ are incompatible 
with net moment  on the Fe, and $\Gamma_3$ is not consistent with the observed 
strong antiferromagnetic coupling between nearest Fe neighbours. 
Hence only the remaining three representations are possible in 
orthoferrites~\cite{doi:10.1063/1.1657530}. 
The $k$$\textendash$search utility in Fullprof was used for obtaining the propagation vector in the case of \tfco\ 
and subsequently, SARA$h$ was used to obtain the magnetic
representations of the allowed magnetic structures. 
After testing the three different magnetic representations along with the nuclear phase
in $Pbnm$, a better visual fit to the experimental data with reasonable agreement factors 
were obtained for $\Gamma_2$, and was accepted as the solution of the magnetic structure 
at 350~K. 
The goodness$\textendash$of$\textendash$fit for the magnetic refinement, $R_\mathrm{mag}$, 
for the three representations are as follows: 
$\Gamma_1$ = 25.4, $\Gamma_4$ = 95.2, $\Gamma_2$ = 17.4.
%
\begin{table*}[!t]
	\setlength{\tabcolsep}{16pt}
	\caption{\label{tab:str} Structural parameters and selected bond distances and bond
		angles of \tfco\ at different temperatures obtained from neutron 
		diffraction. The nuclear space group is $Pbnm$ where the 
		atomic positions are Tb $4e$ ($x,y,z$), Cr/Fe $4b$ (0.5,0,0.5) 
		and O $4e$ ($x,y,z$). Long ($l$) and short ($s$) bond lengths correspond to $M \textendash$O(2) 
		bonds in the $ab$ plane. Medium ($m$) bond length corresponds to the out$\textendash$of$\textendash$plane 
		$M\textendash$ O(1) apical bond.
	}
	\begin{tabular}{l l l l l l l}	\hline	\hline 
		& 350~K & 300~K & 215~K & 100~K & 20~K & 7.7~K  \\  		\hline 
		$a$ (\AA) & 5.3111(4) & 5.3125(6) & 5.3100(6) & 5.3121(0) & 5.3140(5) & 5.3124(2) \\ 
		$b$ (\AA) & 5.5548(1) & 5.5560(5) & 5.5517(5) & 5.5442(0) & 5.5403(4) & 5.5391(3) \\
		$c$ (\AA) & 7.6117(8) & 7.6111(0) & 7.6039(1) & 7.5963(3) & 7.5929(1) & 7.5917(3)\\
		Fe(Cr)$\textendash$O1 ($m$) (\AA) & 1.9885(5) & 1.9902(4) & 1.9886(0) & 1.9866(5) & 1.9858(6) & 1.9850(6)\\
		Fe(Cr)$\textendash$O2 ($l$) (\AA) & 2.0106(0) & 2.0115(4) & 2.0120(3) & 2.0083(8) & 2.0076(0) & 2.007(0) \\
		Fe(Cr)$\textendash$O2 ($s$) (\AA) & 1.9945(5) & 1.9919(8) & 1.9917(4) & 1.9904(0) & 1.9904(2) & 1.9999(5) \\
		Fe(Cr)$\textendash$O1$\textendash$Fe(Cr)($^o$) & 146.2(5) & 145.9(0) & 145.8(5) & 145.8(5) & 145.8(3) & 145.9(2) \\
		Fe(Cr)$\textendash$O2$\textendash$Fe(Cr)($^o$) & 147.2(4) & 147.5(0) & 147.2(3) & 147.1(5) & 147.5(1) & 146.5(4) \\ \hline 	\hline 
	\end{tabular} 
\end{table*}
Figure~\ref{fig:neutron} (a) shows the neutron diffraction patterns at  350~K along with 
the refinement patterns using $Pbnm$ nuclear space group and the magnetic structure according to
$\Gamma_2$ representation. 
The nuclear space group of \tfco\ at all temperatures till 7.7~K was  found 
to be $Pbnm$. 
The refined values of the lattice and bond parameters at different 
temperatures are given in Table~\ref{tab:str}. 
Here, three different $M \textendash O$ bond lengths are listed. 
Long ($l$) and short ($s$) bond lengths correspond to $M \textendash O(2)$ 
bonds in the $ab$ plane while the medium ($m$) bond length corresponds to the out of plane 
$M \textendash O(1)$ apical bond length which is almost parallel to the $c$ axis.   
\\
\begin{figure*}
		\centering
	\includegraphics[scale=0.2]{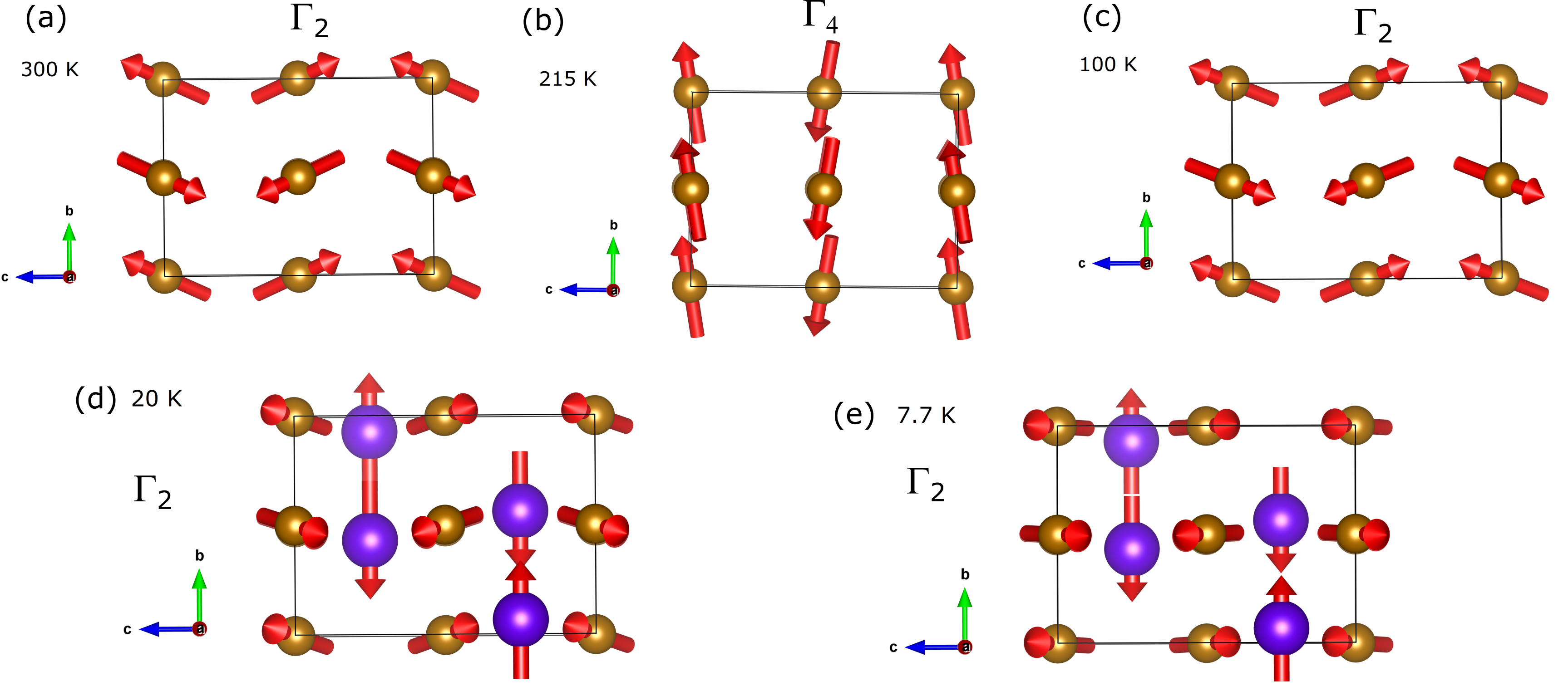}\\
	\caption{\label{fig:magstr} (a, b, c) The magnetic structures of the
		Fe/Cr sublattice at 300~K ($\Gamma_2$), 215~K ($\Gamma_4$) and 
		100~K ($\Gamma_2$) respectively. The $\Gamma_2$ structure 
		remains stable down to 7.7~K, which was the lowest
		probed temperature by neutrons in this study.
	(d and e) The magnetic structures of Tb$^{3+}$ at 20~K and 7.7~K. The Fe/Cr atoms are represented as yellow spheres and Tb as blue}
\end{figure*}
\indent 
As understood from the magnetization data presented in Figure~\ref{fig:mt} (a), 
a magnetic phase transition occurs in \tfco\ at \tn\ = 257~K. 
Refinement of the diffraction pattern suggests that the 
nuclear structure is $Pbnm$ and the magnetic structure is $\Gamma_4$ ($Pb'n'm$) at 215~K. 
Thus, the magnetic structure changes from $\Gamma_2 \rightarrow \Gamma_4$ at \tn. 
The refined neutron diffraction pattern at 215~K is presented
in Figure~\ref{fig:neutron} (b). 
Interestingly, we observe a second spin reorientation transition back to the $\Gamma_2$ ($Pbn'm'$) 
structure at 100~K. 
This temperature is below \tsr\ (190~K) which is identified through the 
derivative of magnetization curve.
The $\Gamma_2$ magnetic structure remains stable down to 7.7~K. 
%
%
In Figure~\ref{fig:neutron} (e), the neutron diffraction intensity of \tfco\ at 
7.7~K is presented after subtracting the contribution of the empty vanadium can
that was used as the sample holder. 
Preliminary evidence for diffuse magnetism is present in difference curves.  
This observation lends support to the short$\textendash$range magnetic fluctuations with a 
close link to the 
features similar to those of Griffiths phase in the magnetization of \tfco.
We attempted to analyze the diffuse intensity by fitting to the Lorentzian curve, 
which is shown as a red solid line in Figure~\ref{fig:neutron} (e). 
The fit enabled us to extract a correlation length of approximately 9~{\AA}. Elastisc and inelastic experiments are underway in single crystals of \tfco\ to study the 
diffuse signatures. 
\\
\indent Further, the rare earth magnetic atoms in $R$Fe$_{0.5}$Cr$_{0.5}$O$_3$ is reported to develop
magnetic ordering at low temperature below 15~K~\cite{bolletta2018spin}.
Our data is in agreement to this observation. 
Our refinement at 7.7~K are consistent with the picture that the Tb$^{3+}$ moments are magnetically ordered
in F$_x$C$_y$ magnetic structure with a ferromagnetic component 
along the $a$ axis. 
It is reported in a recent work~\cite{PhysRevB.98.134417} on \tfco\ that 
only the C$_y$ part remains whereas the ferromagnetic
interactions disappear with the spin reorientation at 1.9~K.
As a result, diffuse magnetic scattering features emerge;
this is well$\textendash$captured in our work as can be seen in
Figure~\ref{fig:neutron} (e).
The magnetic structures of the transition metal and rare earth
moments as a function of temperature are shown in 
Figure~\ref{fig:magstr}.

\subsection{Thermal conductivity and Raman spectroscopy}
\label{thermalcond} 
\begin{figure}[!b]
\centering
\includegraphics[trim={0.2cm, 0, 0, 0}, scale=0.36]{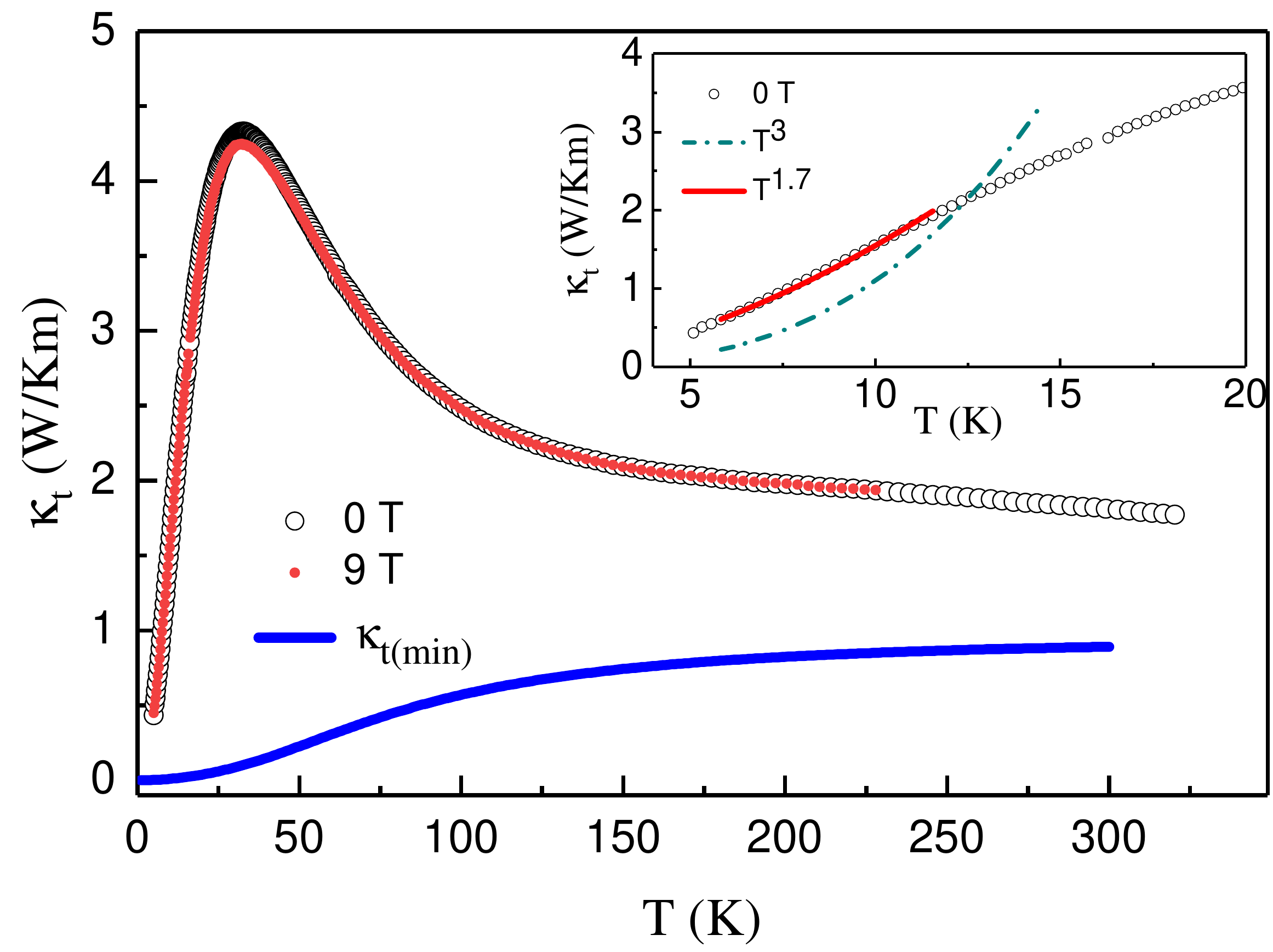}
\caption{\label{fig:thermalcond} Variation of thermal conductivity, $\kappa_\mathrm{t} (T)$
of \tfco\ as a function of temperature. As seen, there is no appreciable change in $\kappa_t (T)$
with the application of 9~T. The magnetic anomalies at \tn\ and \tsr\ seen in the derivative of magnetization are not observed in $\kappa_\mathrm{t} (T)$ or in the derivative (not shown). The blue solid line represents the minimum thermal conductivity (see the text). In the inset, solid and the dashed lines represent $T^{1.7}$ and $T^3$ dependence of $\kappa_\mathrm{t} (T)$ respectively.}
\end{figure}

\begin{figure}[!t]
	\centering
	\includegraphics[trim={0, 0, 0, 0}, scale=0.25]{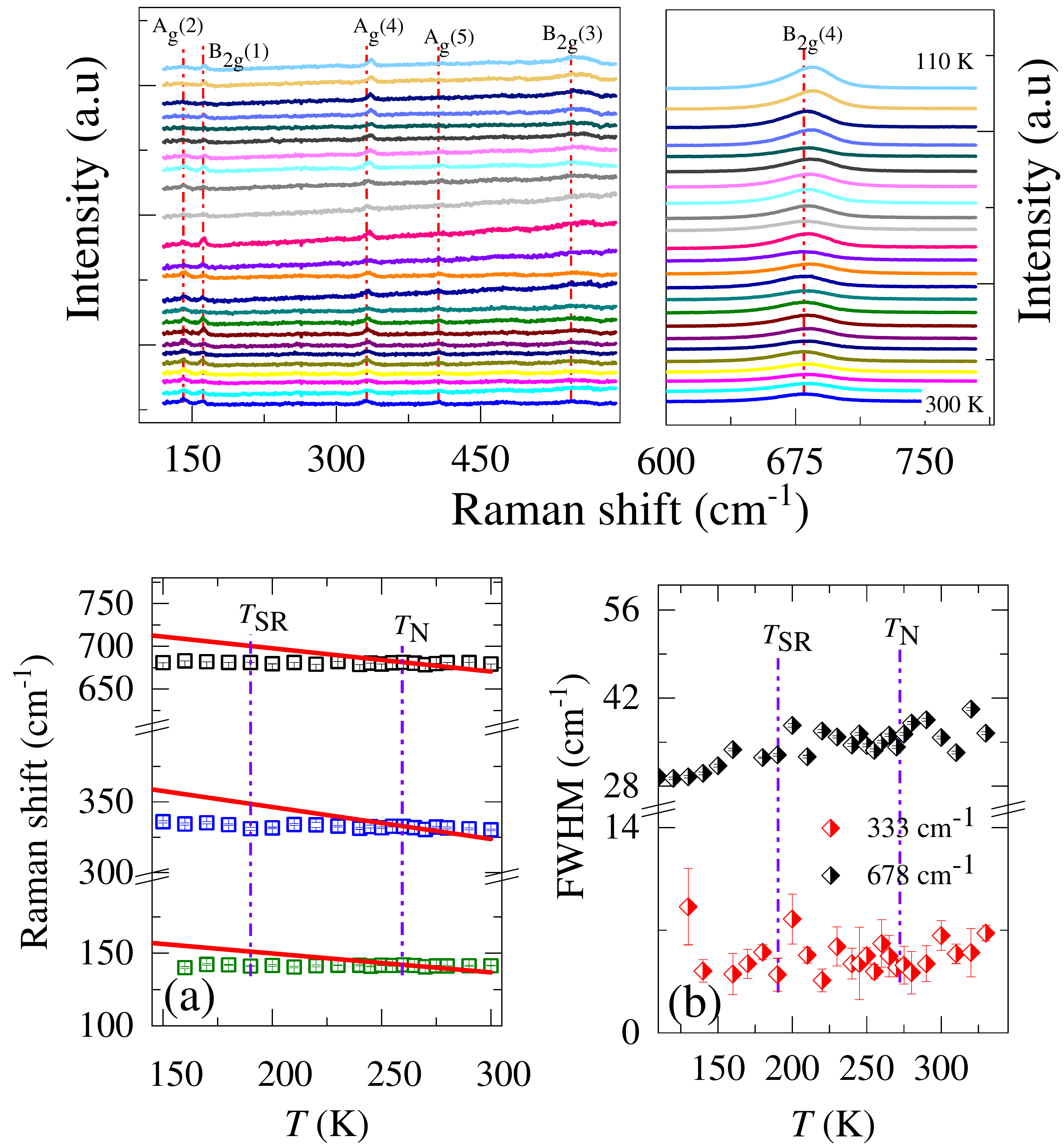}
	\caption{\label{fig:Raman spectrum} Top panel: Raman spectrum of \tfco\ at 
		different temperatures along with the most intense phonon modes assigned. 
		(a) Shows temperature dependence of phonon frequency obtained from the fit of 
		the spectral profile with Lorentzian function. 
		Red solid line shows anharmonic function fitting. 
		(b) Shows temperature dependence of phonon linewidth. 
		Vertical violet dashed line shows position of \tn and \tsr.}
\end{figure}

Figure~\ref{fig:thermalcond} shows the thermal conductivity, $\kappa_t (T)$ of \tfco\ measured in zero and in an applied magnetic field of 9~T. The overall magnitude and temperature dependence of the thermal conductivity suggest that the lattice thermal transport dominates in this material. As can be seen from the figure, there is no appreciable change in $\kappa_t (T)$ with the application of magnetic field. The relatively low value of $\kappa_t (T)$ might point to the presence of disorder in this material. In the context of the presence of the atomic disorder in UZr$^2$ and its impact to heat transport behavior, it is helpful to compare the measured thermal conductivity to the theoretically achievable minimum of the lattice contribution (fully disordered structure). In this model the $\kappa_{t(min)} (T)$ dependence can be calculated by using Debye approximation and assuming that the transverse and longitudinal acoustic phonon modes are indistinguishable~\cite{CAHILL1989927}. The results obtained for \tfco\, using the Debye temperature, $\theta_D$ = 380 K~\cite{VAGADIA20181031} and number of atoms per unit volume, n = 4.7528 m$^{-3}$, are shown in Figure~\ref{fig:thermalcond} by the blue solid line.     
The magnetic anomalies that occur at \tn\ and \tsr\ (seen in the derivative of 
magnetization) are absent in the derivative of $\kappa_t (T)$ (not shown here).
In general, the behavior of $\kappa_t (T)$ of \tfco\ is similar to the thermal 
conductivity variation in other $R$FeO$_3$ compounds like YFeO$_3$, 
GdFeO$_3$ and DyFeO$_3$~\cite{zhao2017comparative}. However, in 
the work by Zhao {\em et al.}~\cite{zhao2017comparative}, single crystal 
samples of orthoferrites were studied in the milli$\textendash$Kelvin temperature range and in 
external magnetic fields up to 14~T. In earlier studies of GdFeO$_3$ and 
DyFeO$_3$, anomalies that occur in the vicinity of the magnetic transitions 
were reflected in the thermal conductivity response also~\cite{zhao2011magnetic, zhao2014ground}. Significantly, low 
$c$ axis thermal conductivity was observed in the case of
YFeO$_3$, GdFeO$_3$ and DyFeO$_3$, considering that the present 
sample is a polycrystalline pellet, in comparison, we observed 
higher values of thermal conductivity in \tfco. The total thermal conductivity could be compared to 
the $T^3$ boundary scattering limit of phonons~\cite{berman1978thermal}. 
In the inset of Figure~\ref{fig:thermalcond}, the temperature dependence of the
$T^3$ form of $\kappa_t (T)$ is shown as a dashed line. The solid line is 
fitted to the $\kappa_t (T) \propto T^n$ expression where $n$ is varied 
as a free parameter. Here, a value of 1.7(3) was obtained for $n$. 
The $\kappa_t (T)$ curve of DyFeO$_3$ shows a weak curvature at low 
temperature (below 3~K) which is attributable to the magnonic 
contribution of Dy spin system~\cite{zhao2014ground}.  Such a 
concave structure is not readily observed in the present case, however, the 
beginning of such a curvature could be discernible near 2~K.
\\
\indent 
Raman spectroscopy was carried out at different temperatures 
in order to understand the phonon behaviour across the magnetic transitions observed in \tfco. 
Raman spectra was recorded from 110~K to 300~K as shown in 
Figure~\ref{fig:Raman spectrum} (top panel) 
with the most intense mode assignment matching with $R$FeO$_3$~\cite{Weber} 
and $R$CrO$_3$~\cite{Srinu_Bhadram_2013}. 
\tfco\ is an orthorhombically distorted perovskite with $Pbnm$ space group symmetry. 
The irreducible representations corresponding to the phonon modes at the Brillouin zone 
center~\cite{Venugopalan} can be defined as, 
$\Gamma$ = 7$A_g$ + 7$B_{1g}$ + 5$B_{2g}$ + 5$B_{3g}$ + 8$A_u$ + 10$B_{1u}$ + 8$B_{2u}$ + 10$B_{3u}$.
Here, $A_g$, $B_{1g}$, $B_{2g}$, $B_{3g}$ are the 
Raman active modes, $B_{1u}$, $B_{2u}$, $B_{3u}$ are the infrared modes, 
and $A_u$ is inactive mode.
Among them, the modes which are above 300~cm$^{-1}$ are related to the vibrations of oxygen, and 
the modes below 300~cm$^{-1}$ are associated with the rare earth ions~\cite{Singh}. 
However, the Raman vibrational modes corresponding to an orthorhombic structure are
$A_g$ + $B_{1g}$ and 2$B_{2g}$ + 2$B_{3g}$, which are symmetric and antisymmetric modes, 
respectively.	
In contrast, $A_g$ + 2$B_{1g}$ + $B_{3g}$, 2$A_g$ + 2$B_{2g}$ + $B_{1g}$ + $B_{3g}$, and 
3$A_g$ + $B_{2g}$ + 3$B_{1g}$ + $B_{2g}$ are associated with the bending modes, rotation 
and tilt mode of the octahedra, and for the rare earth vibrations, respectively~\cite{Iliev}. 
Raman modes generally shifts to low frequency as the temperature increases accompanied by a 
monotonic increase in full$\textendash$width$\textendash$at$\textendash$half$\textendash$maximum (FWHM)~\cite{Nonato}. 
\begin{figure*}[!t]
	\centering
	\includegraphics[scale=0.55]{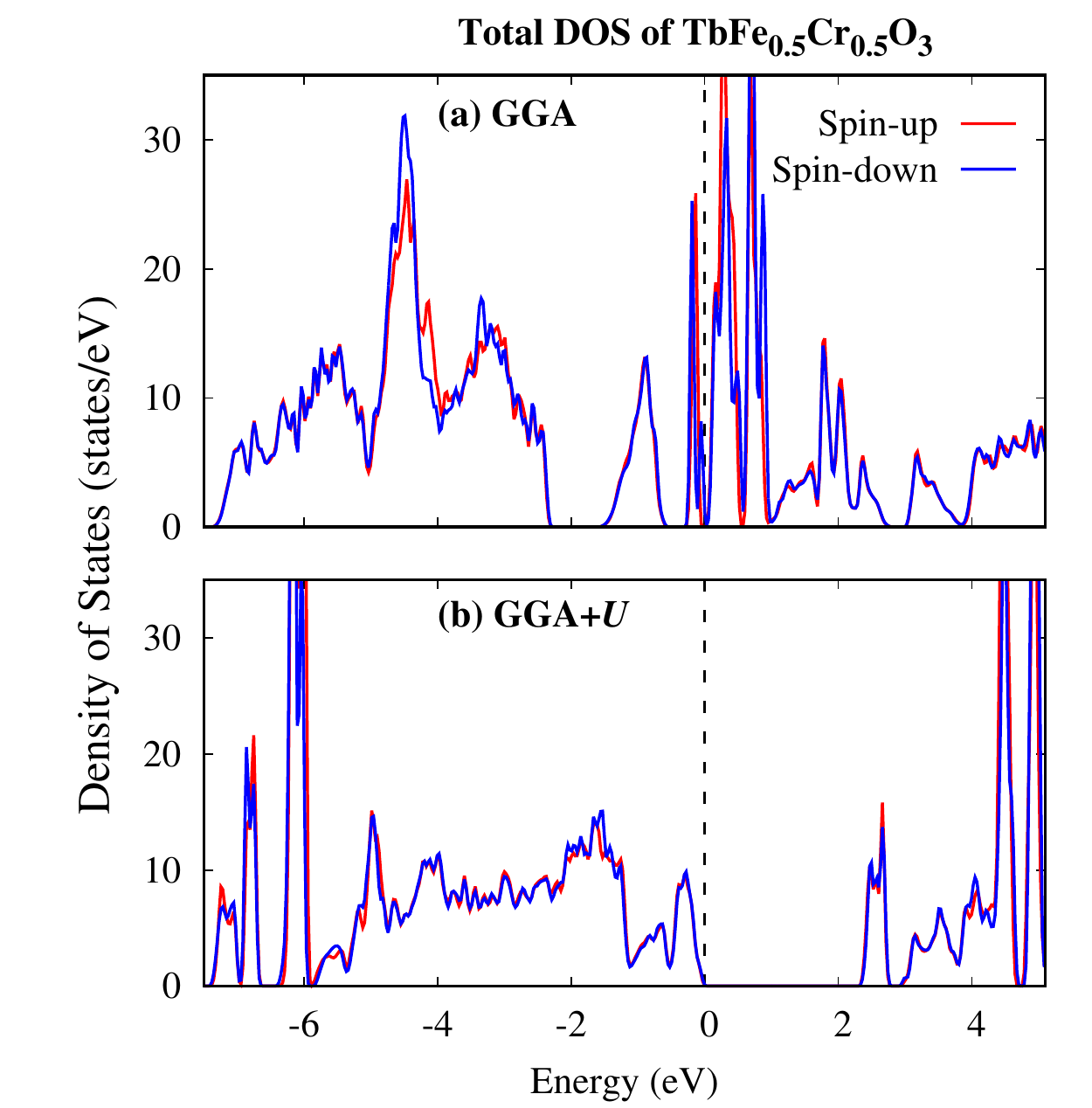}
	\includegraphics[scale=0.4]{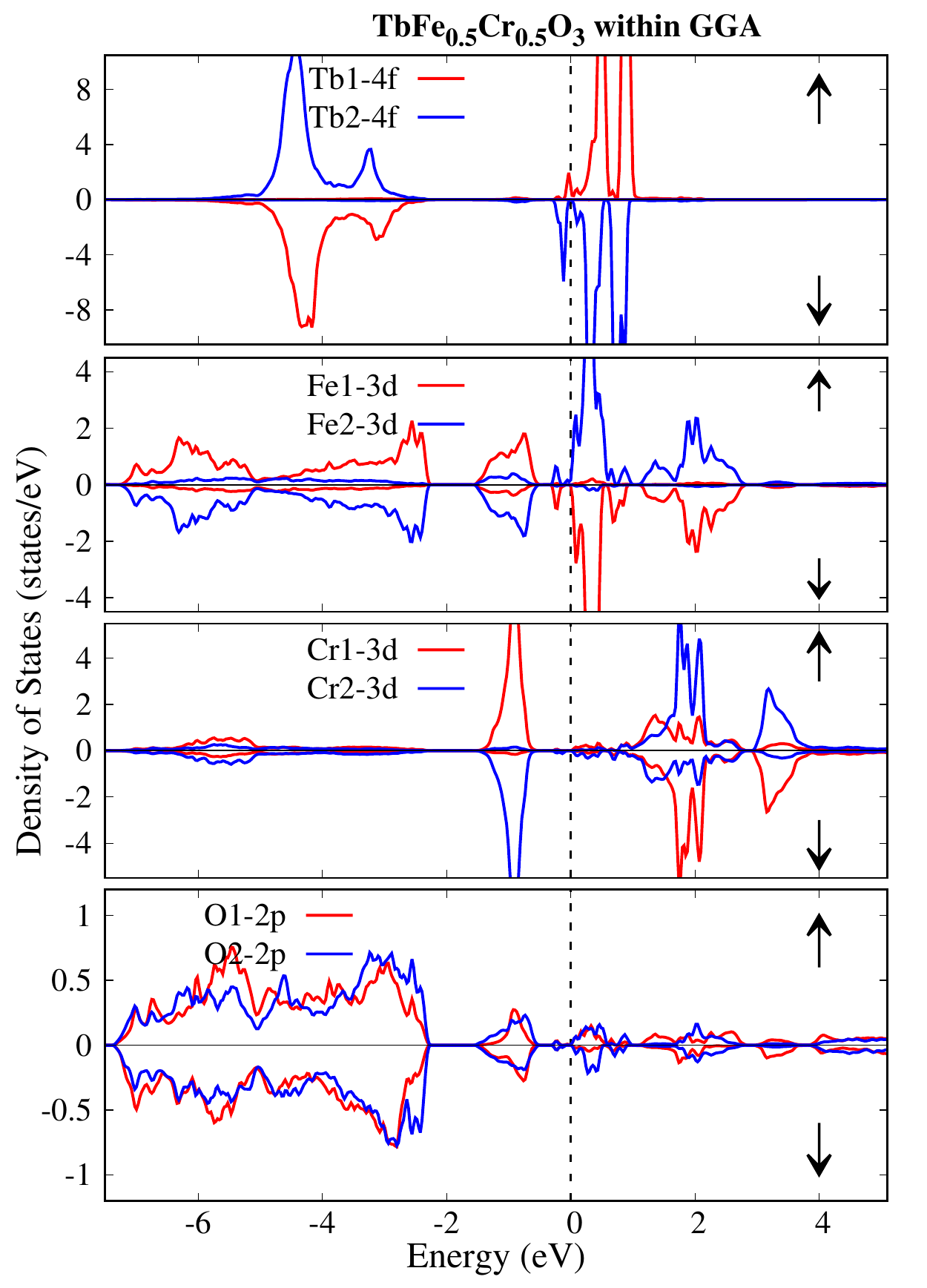}
	\includegraphics[scale=0.4]{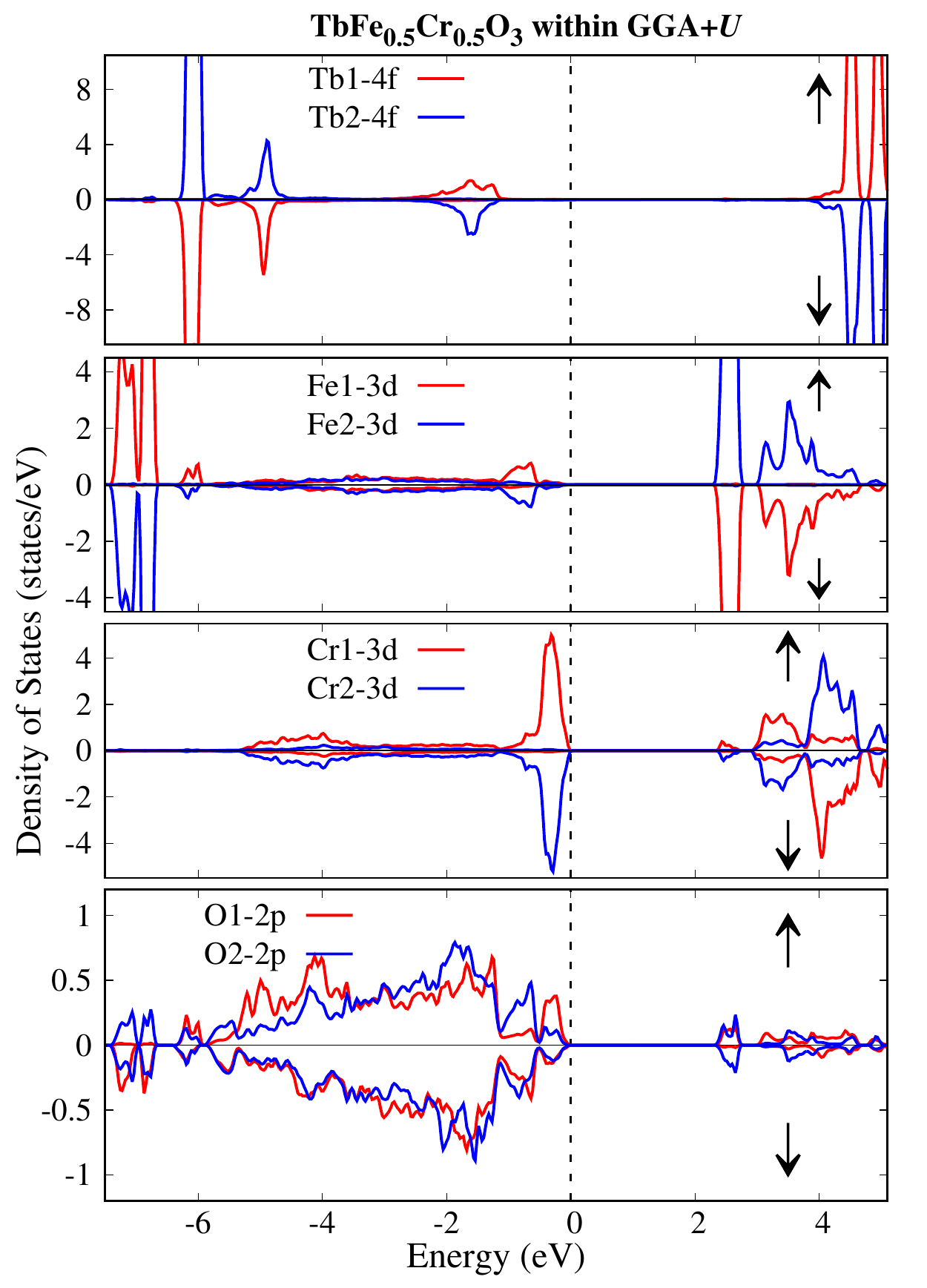}
	\caption{{\label{fig:dos}} Total and partial DOS of TbFe$_{0.5}$Cr$_{0.5}$O$_3$ in AFM1 
		configuration:Total DOS within GGA (top) and GGA+\textit{U} (bottom) (left) functionals; 
		partial DOS contributions from Tb$\textendash$4f, Fe$\textendash$3d, Cr$\textendash$3d and O$\textendash$2p states, respectively within 
		GGA (middle) and GGA+\textit{U} (right) functionals for the spin$\textendash$up and spin$\textendash$down channels.}
\end{figure*}
This is mainly due to the expansion of lattice as thermal energy increases. 
Absence of any extra peak indicates that the spectral symmetry remains same at 
all measured temperature thus confirming that the magnetic transitions are not associated 
with any structural phase transition. 
We analyzed well$\textendash$resolved Raman modes in detail. 
These modes were fitted with Lorentzian function. 
The temperature variation of the phonon frequencies of modes 
$A_g$(2) (141~cm$^{-1}$), $A_g$(4) (333~cm$^{-1}$) and $B_{2g}$(4) (678~cm$^{-1}$) along with the 
fit assuming the standard anharmonic dependence~\cite{Balkanski} of phonon modes are shown in 
Figure~\ref{fig:Raman spectrum} (a).
\\
\indent 
The anharmonic dependence of the modes is given by,
$\omega_\mathrm{anh}$(T) = $\omega_0$$\textendash$C(1 + (2/{($e^{(\hbar\omega/k_BT)}\textendash1)))$}
where $\omega_0$ is temperature$\textendash$independent part of linewidth, C is a constant determined from the fitting, 
$\hbar\omega$ is the phonon energy, and $k_B$ is the Boltzmann constant. 
The sudden change in the phonon frequency near \tn\ and \tsr\ can be clearly seen in Figure~\ref{fig:Raman spectrum} (a). 
A similar kind of anomaly in Raman modes near the magnetic transition was reported in 
$R$CrO$_3$~\cite{Srinu_Bhadram_2013} compounds. 
Magnetostriction can also give rise to similar anomalous behaviour in phonon frequency by modyfying 
unit cell volume~\cite{Nonato}. 
But in that case, FWHM remains unchanged as it corresponds to phonon lifetime which is not 
affected by subtle change in lattice volume caused by magnetostriction. 
But, from Figure~\ref{fig:Raman spectrum} (b), it can be seen that FWHM abruptly drops near 
the magnetic transitions. 
The anomalous change in the mode frequencies and linewidths near the magnetic transition 
establishes the spin$\textendash$phonon coupling in \tfco. 
A similiar signature of spin$\textendash$phonon coupling was reported in $R$CrO$_3$~\cite{Srinu_Bhadram_2013}
and DyFe$_{0.5}$Cr$_{0.5}$O$_3$~\cite{Yin}. 
The possible coupling mechanism involved is the phonon modulation of superexchange integral 
below the magnetic ordering temperature~\cite{Granado}.

\subsection{Density functional theory calculations}
From the total energy calculations for five different collinear magnetic configurations, 
AFM1 ($\uparrow\downarrow\uparrow \downarrow$) is found to be most stable with 
the lowest energy. 
The AFM1 spin structure is found to be consistent with our experimental 
observation for the $\Gamma_2$ state at 7.7~K. 
Similarly, the first excited AFM2 configuration is consistent with the spin 
structure for the $\Gamma_4$ state at 300~K
whose total energy is $\sim 36$ meV per formula unit higher compared 
to the AFM1 state of \tfco.
The order of relative stability of the magnetic states are 
AFM1 $>$ AFM2 $>$ FIM2 $>$ FIM1 $>$ FM. 
This may be an indication of the competing ground state between 
AFM1 and AFM2 observed as a 
Griffiths phase transition from $\Gamma_2$ to $\Gamma_4$ and, 
subsequently, the reentrant to $\Gamma_2$ phase as seen in Figure ~\ref{fig:magstr}.
The magnetic anisotropy energy calculated is $\sim$ 4.68~meV per 
formula unit of \tfco~ with in$\textendash$plane easy axes.
%
In \tfco, the lanthanide Tb takes the charge state $3+$ with 4f$^8$ configuration. 
Likewise, the transition element Fe nominally takes the charge state $3+$ with 3d$^5$
and Cr with charge state $3+$ should take the 3d$^3$ configurations, respectively. 
In the stable AFM1 state, the calculated spin moment at each site of Tb, Fe and Cr 
are $\pm$ 5.9 $\mu_B$, $\pm$ 3.65 $\mu_B$, and $\pm$ 2.36 $\mu_B$, respectively. 
Their respective orbital moments are $\pm 1.03 \mu_B$, $\pm 0.05 \mu_B$, and $\mp 0.034 \mu_B$ respectively.
With GGA+$U$ effects, the spin moment of Tb, Fe and Cr turns out to $\pm 5.97 \mu_B$, $\pm 
4.14 \mu_B$, and $\pm 2.57 \mu_B$ respectively.
The total magnetic moment compensates to zero as Tb, Fe and Cr couples 
antiferromagnetically 
among each other as observed in Figure~\ref{fig:magstr}. 
\\
\indent 
We now proceed to the electronic
structure of \tfco~ in AFM1 state within GGA and GGA+$U$, respectively. 
The spin$\textendash$resolved total and partial density of states (DOS) are shown in Figure~\ref{fig:dos}. \tfco~ is found to be insulating with a band gap of $\sim 0.12$ (2.4) eV within GGA (GGA$+U$). 
The correlation effects $'U'$ significantly changes the electronic behaviour. 
As seen in the partial DOS, the main contributions from Tb$\textendash$$4f$ states that were observed around $E_{\rm F}$ 
are shifting away from each other. 
Those states that are fully occupied shift deep in the valence region while the un-occupied state 
moves far away in the conduction region. 
Similar features were observed also for Fe$\textendash$$3d$ states around $E_{\rm F}$.
On the other hand, Cr$\textendash$$3d$ states are contributing at and around $E_{\rm F}$ hybridizing 
strongly with the O$\textendash$$2p$ orbitals (see partial DOS in Figure~\ref{fig:dos}).  
This is mainly due to the hybridization between the $3d$ states of Cr and Fe with the O$\textendash$$2p$ states.
From the partial DOS contributions of Fe$\textendash$$3d$, three $t_{2g}$ and two $e_g$ are fully occupied by five electrons 
in spin$\textendash$up but in Cr$\textendash$$3d$, three $t_{2g}$ are fully occupied in spin up channel while $e_g$ bands are empty.

\section{Conclusions} 
\indent
We observe an antiferromagnetic transition \tn\ at 257~K and a spin 
reorientation transition \tsr\ at 190~K in the orthoferrite \tfco. 
Interestingly, a reentrant spin reorientation is seen in this compound, 
where the spins reorient again at 100~K.  Through detailed neutron diffraction 
experiments and analysis we find that the spin structure changes from the $\Gamma_2$ representation at
350~K to $\Gamma_4$ at 215~K and then reverts to $\Gamma_2$ at 100~K. 
This structure remains stable until 7.7~K. A clear signature of Griffith phase is 
observed in the magnetization response of \tfco\ and also short$\textendash$range 
spin fluctuations that extend up to high temperature.  
The value of thermal conductivity is low in \tfco\ which is not affected 
by the application of magnetic field of 9~T. The magnetic anomalies at \tn\ and \tsr\ 
are not directly seen in the thermal conductivity data, but
the latter is dominated by the phonon contributions. 
Raman spectroscopic investigation reveals clear evidence of spin$\textendash$phonon coupling in this compound.

\section{Acknowledgments}
The authors acknowledge Center for Nano Science and Engineering (CeNSE), Indian Institute of Science, Bengaluru.
B.M. acknowledges financial support from University Grants commission (UGC), India for Senior research fellowship (SRF).
H.S.N. acknowledges faculty start-up grant from UTEP and Rising Stars award. 
Work at INL was supported by DOE's Early Career Research Program.	
M.P.G. acknowledges the Higher Education Reform Project (HERP DLI-7B) of Tribhuvan University, Kirtipur, Nepal for
the start-up grant, and Alexander von Humboldt Foundation, Germany for the partial support as return fellowship. 
S.R.B. thanks NAST, Nepal for the PhD fellowship, and IFW-Dresden for funding during research stay in Germany. 
M.P.G. and S.R.B. thanks Manuel Richter for the fruitful discussion and Ulrike Nitzsche for the technical assistance.

\bibliographystyle{apsrev}
\bibliography{ref_tfco_v8}

\end{document}